\documentclass[aps,prl,twocolumn,floatfix]{revtex4}
\usepackage{amsmath}
\usepackage{graphicx}

\def\){\right)}
\def\({\left(}
\def\]{\right]}
\def\[{\left[}

\newcommand{\be}{\begin{equation}}
\newcommand{\ee}{\end{equation}}
\newcommand{\gsim}{\, \raisebox{-0.8ex}{$\stackrel{\textstyle >}{\sim}$ }}

\newcommand{\roughly}[1]%
{\mathrel{\raise.4ex\hbox{$#1$\kern-.75em\lower1ex\hbox{$\sim$}}}}

\newcommand\beq{\begin{eqnarray}}\newcommand\eeq{\end{eqnarray}}

\def\Dsl{\,\raise.15ex \hbox{/}\mkern-12.8mu D}

\def\fm3{fm$^{-3}$}

\begin{document}
%
\preprint{\vbox{\hbox{LA-UR-05-1389}}}

\title{Asymmetric Two-component Fermion Systems in Strong Coupling}

\author{J. Carlson and Sanjay Reddy }

\affiliation{Theoretical Division, Los Alamos National Laboratory, Los
Alamos, NM 87545 \\ 
}
\begin{abstract}
We study the phase structure of a dilute two-component Fermi system
with attractive interactions as a function of the coupling and the
polarization or number difference between the two components. In weak
coupling, a finite number asymmetry results in phase separation.  A
mixed phase containing symmetric superfluid matter and an asymmetric
normal phase is favored. With increasing coupling strength, we show
that the stress on the superfluid phase to accommodate a number
asymmetry increases.  Near the infinite-scattering length limit, we
calculate the single-particle excitation spectrum and the ground-state
energy at various polarizations.  A picture of weakly-interacting
quasi-particles emerges for modest polarizations. In this regime near
infinite scattering length, and for modest polarizations, a
homogeneous phase with a finite population of excited quasi-particle
states characterized by a gapless spectrum should be favored over the
phase separated state.  These states may be
realized in cold atom experiments.
\end{abstract}
\maketitle

Recent experiments on cooled Fermi atoms
~\cite{Thomas:2004ex,Bartenstein:2004,Chin:2004,Greiner:2004} and
theoretical developments in dense QCD~\cite{QCDReview} have motivated
renewed theoretical interest in Fermion superfluids.  In Fermi systems
it is well known that attractive interactions destabilize the Fermi
surface. This instability is resolved by the BCS mechanism
characterized by pairing between spin-up and spin-down particles with
opposite momenta on the Fermi surface. The system exhibits superfluid
properties and has a gap in the excitation spectrum.  Measurements
have been performed probing the equation of state (EoS)
~\cite{Bartenstein:2004} and the pairing gap
~\cite{Chin:2004,Greiner:2004} in the strongly-interacting regime.
Quantum Monte Carlo (QMC) methods have also been employed to examine
these properties~\cite{Carlson:2003,Chang:2004,
Astrakharchik:2004,Chang:2005}.

These studies have primarily addressed the unpolarized system. In
contrast, the ground state properties of an asymmetric two-component
system remain unclear as several competing states have been
proposed. These include: a gapless superfluid
~\cite{Sarma:1963,Alford:1999xc,Liu:2002gi,Shovkovy:2003uu,Alford:2003fq};
a mixed phase consisting of BCS and normal components
~\cite{Bedaque:2003hi}; and the Larkin, Ovchinnikov, Ferrel and Fulde
(LOFF) phase where the order parameter acquires a spatial variation
~\cite{Fulde:1965,Larkin:1965} which in three dimensions may manifest
as crystalline state~\cite{Alford:2000ze}. The competition between these states 
is especially relevant to understanding the phase structure of dense quark matter~\cite{QCDReview}. In this letter we address
several issues relating to the ground state of an asymmetric
two-component Fermi system in the strong coupling regime. Using both
mean-field theory and QMC we find that in strong coupling and at
finite polarization the stress on the BCS state to accommodate the
number asymmetry increases. In the intriguing strong-coupling regime
near $k_{\rm F}a=\infty$, a superfluid state with non-trivial gapless
excitations may be favored. We also find that these quasi-particles
are weakly interacting.

{\it Two-component Fermi System:} We consider a system consisting of
non-relativistic spin-up and spin-down Fermions at finite
polarization.  A short-range potential (range $\ll$ inter-particle
distance) between spin-up and spin-down particles characterizes the
interaction. The interaction between spin-up particles (or spin-down
particles) is considered to be negligible compared to the interaction
between up and down spins. We study both weak coupling and strong
coupling limits to explore the phase structure from the BCS-like
regime characterized by pairing at the Fermi surface to the Bose
Einstein condensate regime characterized by bound bosonic states.

The Hamiltonian (Grand-Canonical) is given by:
\begin{align} 
H=\sum_{k,s=\uparrow,\downarrow}  \left( \frac{k^2}{2m} -
\mu_{s}\right)a_{s}^\dagger a_{s} + g\sum_{k,p,q} a_{k+q\uparrow}^\dagger a_{p-q\downarrow}^\dagger
a_{k\uparrow} a_{p\downarrow}\, \nonumber
\end{align}
where $g$ is the effective four-fermion interaction whose strength at
low energy is determined by the two-body scattering length $a$.  For
the two component system, the spin-up and and spin-down chemical
potential may be written as $\mu_{\uparrow} = \mu + \delta \mu$ and
$\mu_{\downarrow} = \mu - \delta \mu$, respectively. The density
$n=n_{\uparrow}+n_{\downarrow}$ determines $\mu$ and the
polarization density $\delta n=n_{\uparrow}-n_{\downarrow}$ determines
$\delta \mu$.  In trapped atom experiments this regime can be 
accessed by adjusting the population of the two species.

Initially, we consider a Fermi system in the ``universal'' regime
characterized by an infinite scattering length. When $a=\infty$, the
interaction does not present a dimensionful scale. Consequently, the
energy density, pressure and chemical potential of the unpolarized
system are related to that of the Fermi gas by the relation
$\epsilon(a=\infty)=\xi ~\epsilon_{\mathrm FG}$, $P(a=\infty)=\xi~P_{\mathrm FG}$, and $\mu = \xi ~k_F^2/2m$,
respectively.  Quantum Monte Carlo studies in small systems with 12-20
particles have determined this numerical coefficient to be
$\xi=0.44(2)$~\cite{Chang:2004}.  The system also exhibits a gap in
the excitation spectrum, numerical studies indicate that the gap
$\Delta= 0.95(5) E_{FG} = \beta \mu$ with $\beta \approx 1.4$.  Below
we present new calculations for larger system sizes, which are useful
to more fully explore the dispersion of the single-particle
excitations.  For these larger systems we find $\xi=0.42(1)$ and
$\Delta = 0.84(5) E_{FG}$.  We note that the value for $\xi$ is in
agreement with experimental studies described in Ref.~\cite{Bartenstein:2004}.

Our primary interest here is the spin-polarized system.  Earlier work
by Bedaque, Caldas and Rupak based on BCS mean-field theory showed
that a finite polarization would lead to phase
separation~\cite{Bedaque:2003hi}. A heterogeneous mixed phase
consisting of an unpolarized superfluid state coexisting with a
partially polarized Fermi gas state was shown to have lower Free
energy than the homogeneous gapless superfluid phase (also called the
breached-pair phase) suggested earlier by Liu and
Wilczek~\cite{Liu:2002gi}.  More recently, Forbes, Gubankova, Liu, and
Wilczek~\cite{Forbes:2004cr} have shown that the this phase may be
stabilized by finite range interactions and different masses for the
two species.  Here, though, we consider the equal mass case with
short-range interactions, examining the phase structure at stronger
coupling.

{\it Normal-Superfluid Mixed Phase:} Polarizing the superfluid state
is disfavored due to the presence of a gap in its excitation spectrum
A heterogeneous state containing normal and superfluid phases provides
an alternate route to accommodate a finite polarization. Here the
excess spin-up particles could reside in the normal phase.

Phase coexistence is possible between states separated by a first
order transition if at fixed chemical potential they can have the same
pressure. Hence we require
\be 
P_{\mathrm{Superfluid}}(\mu,\delta\mu)=P_{\mathrm{Normal}}(\mu,\delta\mu) \,.
\label{gibbs}
\ee
Pressure equilibrium uniquely determines $\delta\mu$ for a given
$\mu$.  We emphasize that at fixed $\mu$, $\delta\mu$ in the mixed
phase does not change with polarization - it is driven to lie exactly
at the first order transition point satisfying Eq.~\ref{gibbs}. An
increase in net polarization is accommodated by an increase in the
volume fraction of the normal phase.

We observe that $\delta\mu$ is the energy required to introduce a
spin-up particle into the normal component of the mixed phase and
$\Delta$ is the corresponding energy in the superfluid component
(where the energy is measured with respect to the chemical potential
$\mu$). In the weak coupling BCS regime, the mixed phase is
characterized by $\delta\mu/\Delta=1/\sqrt{2}$ and the BCS state
remains unpolarized.  When $\delta \mu \ge \Delta$ the BCS state will
acquire a finite polarization resulting in gapless excitations in its
spectrum. Otherwise, a finite polarization will result in phase
separation~\cite{Bedaque:2003hi,Cohen:2005ea}. Thus, at infinitesimal
polarization, $\delta\mu-\Delta$ is the energy difference per unit
polarization between the homogeneous gapless phase and the mixed
state. These observations motivate us to examine how the ratio
$\delta\mu/\Delta$ changes with coupling strength.

We first analyze the situation at $k_{\rm F} a=\infty$. Using the EoS
results of Ref.~\cite{Chang:2004} and conditions of Gibbs equilibrium
(Eq.~\ref{gibbs}) we find that the ratio
\be
\frac{\delta\mu}{\Delta}=\frac{1}{\beta}~
\left(\frac{2^{2/5}}{\xi^{3/5}}-1\right)\,.
\label{dmu}
\ee
Substituting the QMC results for $\xi$ and $\beta$, we find that
$\delta\mu/\Delta=1.00(5)$.  This suggests that the BCS state is close
to the polarization threshold; and that by tuning the coupling
strength cold-atom experiments in traps containing two-Fermion species
should be able to explore the large $\delta\mu/~\Delta$
regime. Further, we find that $\delta\mu \ge \mu$ - indicating that
the polarization in the normal phase is maximal. It is intriguing that
we find $\Delta \simeq \delta\mu$ at $k_{\rm F}a=\infty$.   
\begin{figure}[ht]
\includegraphics[width=\columnwidth]{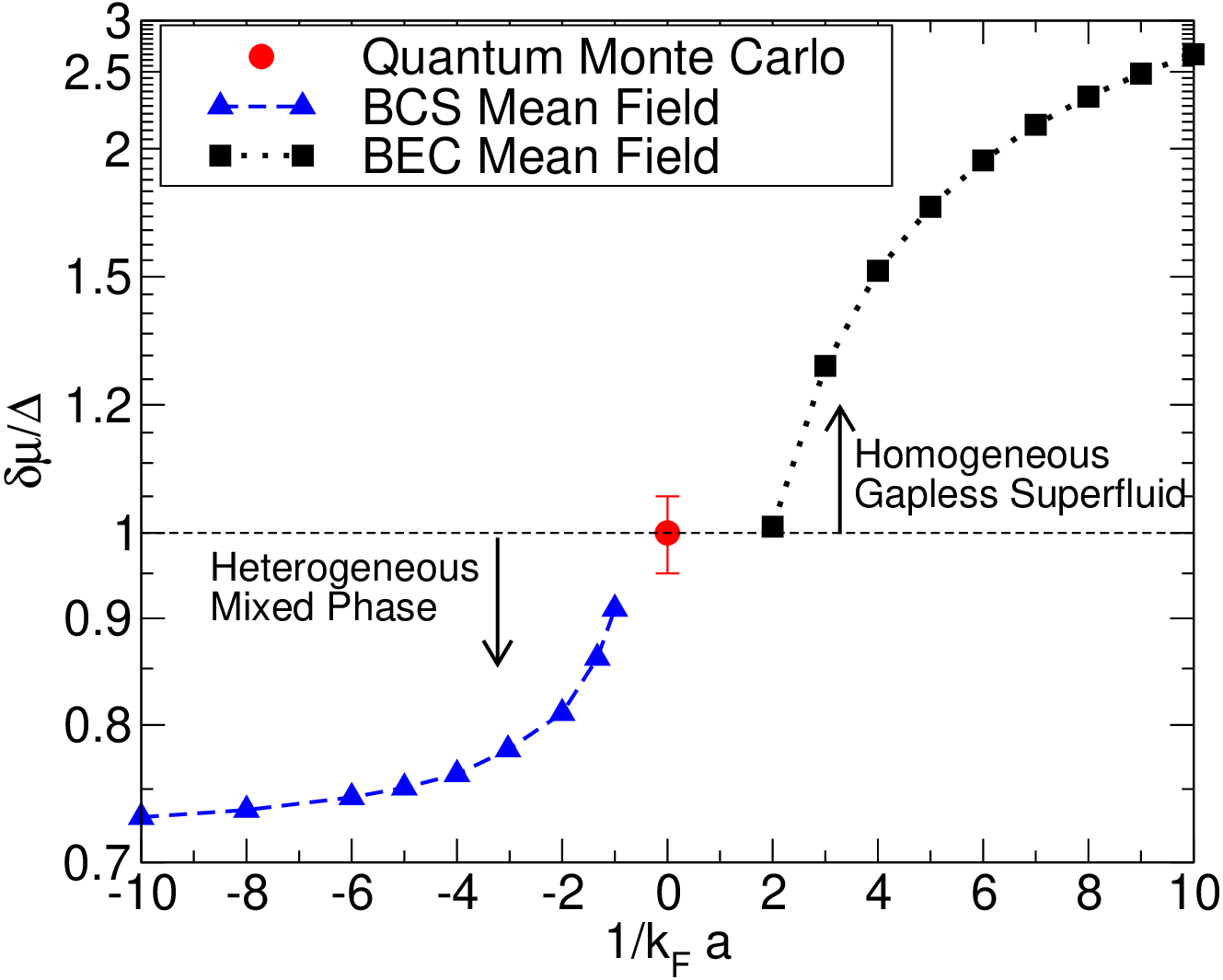}
\caption{Variation of $\delta\mu/ \Delta$ with increasing coupling.}
\label{polar}
\end{figure}

The gap increases with coupling but $\delta\mu$ increases at a faster
rate.  The increase in $\delta\mu$ is driven by the pressure
equilibrium condition. This trend can be demonstrated by using mean
field theory in the weak coupling regime where $\Delta(k_{\rm F}a)=8
\mu \exp{\left(-2+(\pi/2\sqrt{2}k_{\rm F}a)\right)}$. Earlier studies
of the mixed phase ignored the leading order $k_F a$ corrections to
the pressure and chemical potential~\cite{Bedaque:2003hi}. However,
including these corrections is straightforward~\cite{Fetter:1971}.  We
find the pressure in the normal and BCS state is given by
\begin{eqnarray}
P_{\rm Normal}&=&\frac{k_{F\uparrow}^5}{30 \pi^2 m} +
  \frac{k_{F\downarrow}^5}{30 \pi^2 m} + \frac{a}{9 \pi^3 m}
  ~k_{F\uparrow}^3 k_{F\downarrow}^3 \\ 
P_{\rm BCS}&=&\frac{k_{\rm
  F}^5}{ 15 \pi^2 m} + \frac{a}{9 \pi^3 m} ~k_{\rm F}^6 + P_{\Delta}\,
\label{pweak}
\end{eqnarray}
respectively, where $P_{\Delta}=mk_{\rm F} \Delta^2/(4\pi^2)$ is the
pairing contribution to the pressure. The chemical potentials also
receive corrections at order $k_{\rm F}a$. In the normal phase $
\mu_{\uparrow(\downarrow)} = k_{\rm F\uparrow(\downarrow)}^2/2 m + (4 \pi
a/m)~ k_{\rm F \downarrow(\uparrow)}^3/(6 \pi^2)$.

Gibbs equilibrium condition determines the variation
of $\delta \mu$ as a function of $1/k_{\rm F} a$ at fixed $\mu$. While
we expect the mean field result to be valid only for small coupling,
the results which are shown in Fig.~\ref{polar} show the expected
trend - an increase in $\delta\mu/\Delta$ with increasing coupling.
The QMC result at the universal point
$k_{\rm F}a=\infty$ confirms this increase. The behavior in the extreme BEC limit is also shown. In this
region, we may use mean field theory to calculate the self energy of a
spin-up Fermion in the BEC phase. This is given by $\Delta_{\rm BEC}=4
\pi a_{\rm BF}n_{\rm B}/\tilde{m}$, where $a_{\rm BF}\simeq 1.2 a$ is
the Fermion-Boson scattering length~\cite{Petrov:2004}, $n_{\rm B}$ is
the Boson density and $\tilde{m}$ is the reduced mass of the
Fermion-Boson system.  The Boson-Boson scattering length is also known
and is given by $a_{\rm BB} \simeq 0.6 a$~\cite{Petrov:2004}. We use
this to calculate the pressure of the BEC at leading order in the
$n_{B}^{1/3} a_{\rm BB}$ expansion~\cite{Fetter:1971}. Pressure and
chemical equilibrium then uniquely determine the chemical potential of
spin-up Fermions in the Fermi gas phase. In agreement with earlier
work by Viverit, Pethick and Smith~\cite{Viverit:2000} we find that
the homogeneous BEC phase easily accommodates a finite polarization in
the dilute regime - as evidenced by $\delta\mu/\Delta_{\rm BEC} \gg
1$.

\begin{figure}[ht]
\includegraphics[width=\columnwidth]{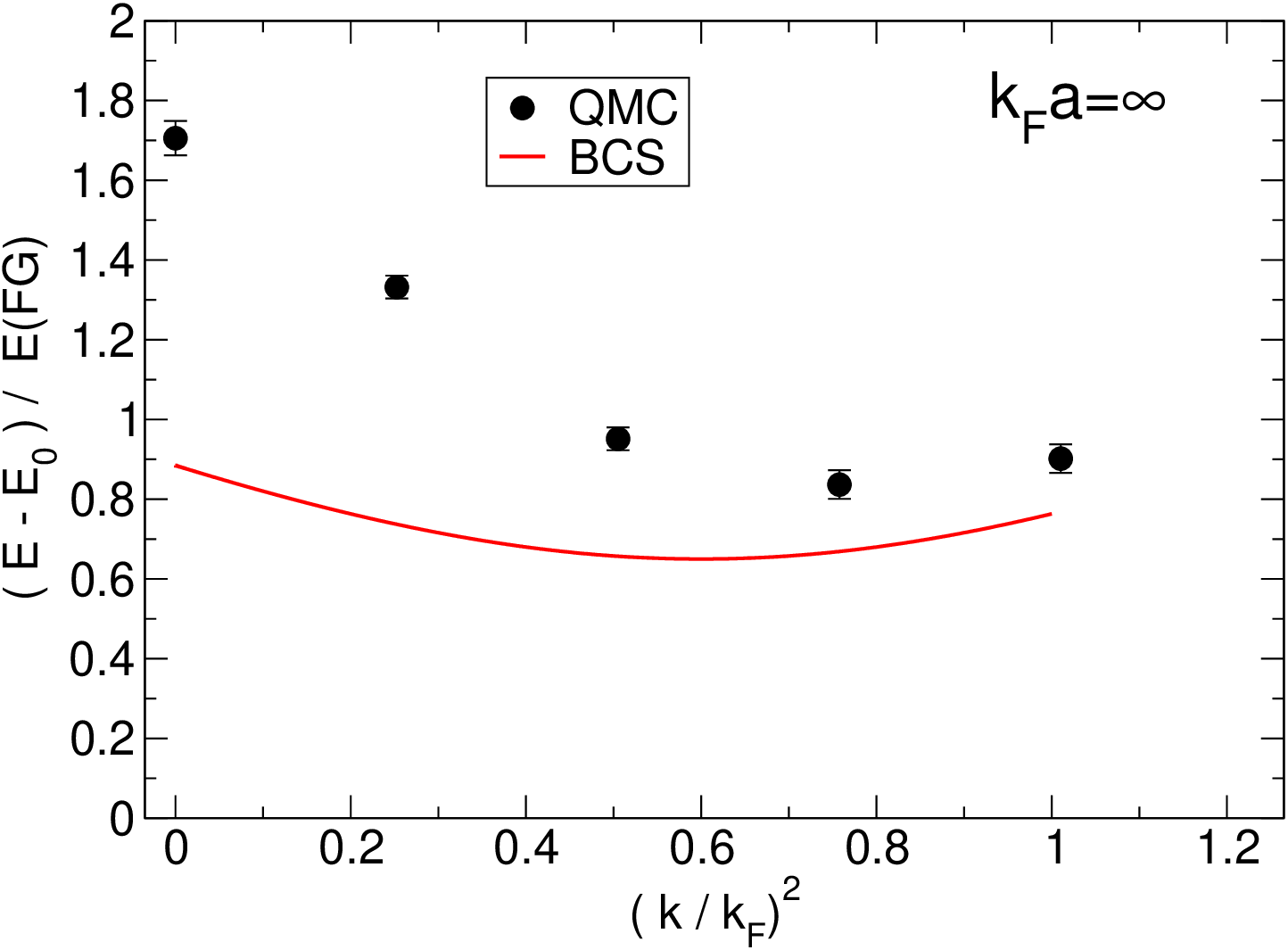}
\caption{The quasi-particle spectrum above the superfluid phase at
$k_{\rm F}a=\infty$. For reference, the quasi-particle spectrum in BCS theory for  $k_{\rm F}a=\infty$ is also shown.}
\label{fig:qps}
\end{figure}
 
{\it Quasiparticle Dispersion and Polarized Superfluid State:} In
addition to calculating the paired ground-state energy and the
superfluid gap, we have performed QMC calculations of the homogeneous
superfluid phase to examine the quasiparticle dispersion as a function
of momentum.  In addition, we have calculated the ground-state energy
of systems at finite polarization to examine the interaction between
these quasi-particles and determine if the gapless phase can support a
macroscopic polarization.  The methods used are identical to those
employed earlier and the same (finite-range) cosh potential was used
~\cite{Carlson:2003,Chang:2004}. As before, we expect the small but
finite-range of the potentail to have a small effect upon the
ground-state energy and the gap. Here we employ somewhat larger system
sizes, however, ranging from 54-66 particles rather than the earlier
studies from 12 to 20 particles.  The larger system sizes allow for a
somewhat finer momentum grid which is useful for examining the
dispersion.  

For the dispersion calculations, we place an unpaired spin in a state of
definite momenta $k = (n_x, n_y, n_z) 2 \pi / L$, where L is the
(cubic) box length and the $n_i$ are integers describing the momenta
in each coordinate.  The BCS plus unpaired particle wave function
can be calculated efficiently as a determinant~\cite{Carlson:2003}.
As for the unpolarized system, we employ the fixed-node 
algorithm to avoid the fermion sign problem.  This yields an upper
bound to the energy for this system, the parameters in the BCS
pair function $\phi (r_{ij})$ are (approximately) optimized to 
yield the best fixed-node energy.  We include Jastrow correlations
between anti-parallel and parallel spins in the trial wave function
to minimize the statistical errors~\cite{Chang:2005}.  These do not
affect the energy, however, as they do not change the nodal surfaces
where $\Psi_T = 0$.

The quasiparticle spectrum at $k_F a = \infty$ is displayed in Figure
\ref{fig:qps}.  The BCS prediction at $k_F a = \infty$ is shown for comparison.
The QMC points are calculated by computing $E_k (N+1)
- [E_0(N)+E_0(N+2)]/2$ at constant density, 
where $E_k (N+1)$ is the energy of the state
of momentum k with 1 unpaired and N paired particles. The $E_0$ are
the ground states of the $N$ and $N+2$ particle systems.

Note that the minimum is at a momentum significantly
less than the Fermi momentum.  For larger coupling, the minimum
in the dispersion will continue to trend towards lower momenta.
This is apparent in the figure from the two sets of QMC
results.  From the QMC calculations we extract
$\xi = 0.42(1)$ and a gap of 0.84(4)$E_{FG}$, or  $\beta = 1.2$,
somewhat smaller than earlier results.  

In order to understand if the BCS phase can support a finite
polarization, it is important to also study the interaction
between the (polarized) quasi-particles.  We have used QMC
techniques to search for the ground state energy as a function
of polarization.  The lowest variational energy in each case is
found by filling the states at the minima of the quasiparticle spectra.

\begin{figure}[ht]
\includegraphics[width=\columnwidth]{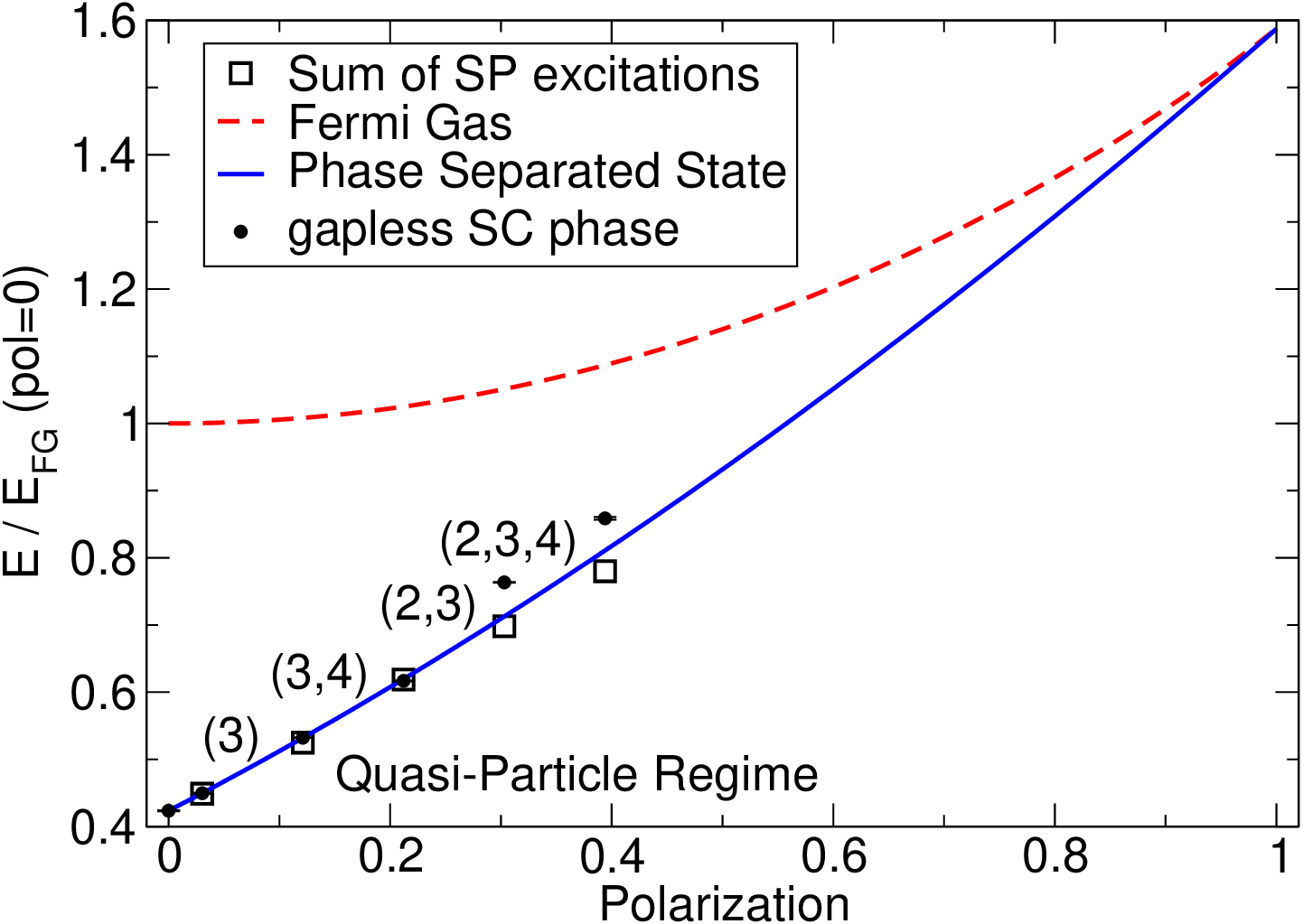}
\caption{Ground-state energy vs. polarization at fixed density and
$a=\infty$. 
}
\label{fig:pol}
\end{figure}

The results are shown in Fig.~\ref{fig:pol}.  These results
demonstrate that the quasi-particles at small polarization are nearly
non-interacting.  This would be expected at small polarizations since
the pair size is expected to be of the order of the inter-particle
separation. The solid points are the QMC calculations with finite
polarization. The integers next to these solid points indicate the
momentum shells $n^2$, where $k^2 = n^2 (2 \pi / L)^2 = ( n_x^2 +
n_y^2 + n_z^2 ) (2 \pi / L )^2$, filled in the trial wave function.
Calculations were performed for 66 particles at various polarizations.

The open symbols represent the sum of single-particle energies
obtained from the single-particle dispersion calculations of Figure
\ref{fig:qps}.  At small polarizations this is nearly degenerate with
the full calculation.  The solid line in the figure is the energy of a
phase separated unpolarized paired phase and a fully polarized Fermi
Gas.  

The calculations indicate that the mixed and homogeneous phases are
essentially degenerate at $k_F a = \infty$.  Our calculations
reproduce momentum distributions similar to those proposed for the
gapless SC state, the states near the minimum of the dispersion are
nearly fully occupied for one species, and nearly unoccupied for the
other.

For larger couplings ($k_F a > 0$) at small polarizations, a
transition to the homogeneous phase is expected to occur at some point
from examining the behavior at very large couplings. Here the system
can be thought of as a dilute mixed Fermi-Bose system where the bosons
are tightly bound pairs of Fermions.  The zero-temperature phase
diagram of such mixtures has been investigated by Viverit, Pethick,
and Smith~\cite{Viverit:2000}.  They find that at small fermion
densities the system is homogeneous, while at larger densities it
phase separates into either a pure fermion plus mixed or pure fermion
plus pure boson phase.

The results shown in Figure \ref{fig:pol} are qualitatively similar.  Beyond
a certain value of polarization, around 0.3 in the figure, the homogeneous
system is clearly higher in energy than a 
phase separated state consisting of an unpolarized BCS 
plus fully polarized fermions.  In this regime it appears
that the BCS component would continue to support a smaller polarization.

{\it Conclusions:} From our calculations it appears that at couplings
near and beyond $k_F a = \infty$ a homogeneous superfluid state with
finite polarization and non-trivial gapless excitations may be
accessible experimentally. We also find that at low polarization, the
polarization is carried by quasi-particles that are nearly
non-interacting and occupy momenta below $k_F$. In cold atom
experiments we expect the trap potential to favor the phase separated
state, while finite-range effects, which are expected to suppress the
gap, and surface energy cost may help stabilize the homogeneous phase
relative to the phase separated state.  A naive interpretation of our
results would indicate that a gapless superfluid exists only at very
strong coupling where $\Delta/\mu \gsim 1$. However in more complex
systems such as dense quark matter, charge neutrality imposed by
long-range forces plays an important role in this phase
competition~\cite{Shovkovy:2003uu,Alford:2003fq,Reddy:2005}. We also
note that we have not analyzed other exotic possibilities such as the
LOFF state which may be relevant.  We are planning on examining such
possibilities in future calculations, and also examining systems at
large polarizations.

{\it Acknowledgments:} We thank Tanmoy Bhattacharya and Eddy
Timmermans for valuable conversations and Mark Alford for detailed
comments on the manuscript.  We would also like to thank the Open
Supercomputing Initiative at Los Alamos National Laboratory and the
National Energy Research Supercomputing Center for providing the
computational resources necessary for these studies.  This research
was supported by the Dept. of Energy under contract W-7405-ENG-36.
\vspace{-0.2in}

\end{document}